\documentclass[twocolumn,aps,prb,floats,longbibliography]{revtex4-1}
\usepackage[latin9]{inputenc}
\usepackage{epsfig}
\usepackage{graphicx}
\usepackage[figuresright]{rotating}

\usepackage{amsfonts}
\usepackage{amsmath}
\usepackage{amssymb}
\usepackage{wasysym}
\PassOptionsToPackage{normalem}{ulem}
\usepackage{ulem}
\usepackage[unicode=true,pdfusetitle,
 bookmarks=true,bookmarksnumbered=false,bookmarksopen=false,
 breaklinks=false,pdfborder={0 0 1},backref=false,colorlinks=false]
 {hyperref}
\hypersetup{colorlinks,linkcolor=blue,citecolor=blue,urlcolor=blue}

\makeatletter

\usepackage{xcolor}
\usepackage{pdfcolmk}
\usepackage{wasysym}

\usepackage{txfonts}

\def\be{\begin{equation}} \def\ee{\end{equation}}
\def\bea{\begin{eqnarray}} \def\eea{\end{eqnarray}}

\makeatother

\begin{document}
\title{Distinguishing thermodynamics and spectroscopy for octupolar U(1) spin liquid of Ce-pyrochlores} 
\author{Gang Chen}
\affiliation{International Center for Quantum Materials, School of Physics, Peking University, Beijing 100871, China}
\affiliation{Department of Physics and HKU-UCAS Joint Institute for Theoretical and Computational Physics at Hong Kong, 
The University of Hong Kong, Hong Kong, China}
\affiliation{Collaborative Innovation Center of Quantum Matter, Beijing 100871, China}

\date{\today}
\begin{abstract}
Inspired by the progress on the spin liquid candidates Ce$_2$Sn$_2$O$_7$ 
and Ce$_2$Zr$_2$O$_7$ where the Ce ions carry the dipole-octupole doublets, 
we analyze the distinction between the thermodynamic and spectroscopic 
measurements for the octupolar U(1) spin liquid. Due to the peculiar properties 
of octupolar U(1) spin liquid and the selective Zeeman coupling, 
both uniform susceptibility and the spectroscopic measurements only probe 
the spinon excitations, while the specific heat measures all excitations.   
After clarifying the contents in each measurement, we observe that,  
thermodynamics carries the information of single spinon, 
while the spectroscopy is associated with spinon pairs. 
Such distinction immediately leads to the multiplicity relation of the excitation 
gaps from different measurements that include the uniform susceptibility, 
Knight shift and $1/T_1$ spin lattice relaxation of NMR and $\mu$SR, 
and the inelastic neutron scattering measurements. 
We hope this work provides a useful recipe for the identification 
of fractionalization in the octupolar 
U(1) spin liquid and other spin liquids. 
\end{abstract}

\maketitle

\section{Introduction}
\label{sec1}

Deconfinement and fractionalization have played an important role 
in our understanding of elementary excitations in the deconfined phases 
such as the one-dimensional spin chains and the high-dimensional 
topological states and spin liquids~\cite{Savary_2016,Knolle_2019}. 
In weakly-coupled spin chains with spin-1/2 local moments, 
the deconfined spinons can propagate 
within the chains and can only propagate in pairs between chains~\cite{Kohno_2007}. 
In high-dimensional deconfined phases such as topological states and spin liquids,
the fractionalized excitations are deconfined in all spatial dimensions~\cite{Savary_2016,Knolle_2019},
and thus fundamentally modify the thermodynamic and dynamic properties 
of the systems. The detection and manipulation of these fractionalized   
and deconfined quasiparticles has become one of the central questions 
in modern condensed matter physics. In this work, we attempt to make use of 
the bare property of deconfinement to analyze the thermodynamics and spectroscopic 
properties for the pyrochlore U(1) spin liquid~\cite{PhysRevB.69.064404}. In particular, 
we focus on the octupolar U(1) spin liquid in the pyrochlore magnets 
with dipole-octupole local moments~\cite{PhysRevLett.112.167203,PhysRevB.95.041106}. 
We hope to find sharp experimental evidences simply based on the deconfinement and fractionalization 
without invoking extra ingredients.

The rare-earth pyrochlore magnets~\cite{PhysRevLett.112.167203,PhysRevLett.115.097202,PhysRevB.95.041106,PhysRevResearch.2.013334,Gao_2019,Gao_2022,Sibille_2020,PhysRevLett.122.187201,PhysRevX.12.021015,PhysRevLett.124.097203,petit2016observation} 
with the Nd$^{3+}$ and Ce$^{3+}$ ions and 
the rare-earth spinel magnets~\cite{PhysRevB.99.134438} 
such as MgEr$_2$Se$_4$ 
have been found to be the hosting materials for the dipole-octupole (DO) doublets. 
In particular, Ce$_2$Sn$_2$O$_7$ and Ce$_2$Zr$_2$O$_7$ 
exhibit the spin liquid behaviors~\cite{PhysRevLett.115.097202,Gao_2019,Sibille_2020,PhysRevLett.122.187201,PhysRevX.12.021015,Gao_2022}, 
and more recent investigation 
has been applied to Ce$_2$Hf$_2$O$_7$~\cite{PhysRevMaterials.6.044406}. 
The DO doublet is a special Kramers doublet where each state of the doublet 
is a singlet representation of the $D_{3d}$ point group~\cite{PhysRevLett.112.167203}.  
In the specific case of Ce$_2$Sn$_2$O$_7$ and Ce$_2$Zr$_2$O$_7$~\cite{PhysRevLett.115.097202,Gao_2019,Sibille_2020,PhysRevLett.122.187201,PhysRevX.12.021015,Gao_2022},
although the Ce$^{3+}$ ion has a total ${J=5/2}$, the ground state doublet 
of the crystal field has the wavefunction ${|{J^z = \pm \frac{3}{2} } \rangle}$ 
that is an integer multiple of $ \frac{3}{2} $ for $J^z$ and 
is thus a DO doublet where $J^z$ is defined on the local $111$ 
axis of each pyrochlore sublattice and will be omitted in the local state
expression below. The effective spin-${1}/{2}$ moment, 
${\boldsymbol S}$, is then defined on the ground state doublet with 
${S^z = \frac{1}{2} [ | \frac{3}{2}\rangle  \langle \frac{3}{2}| -  | {- \frac{3}{2}}\rangle\langle  - \frac{3}{2}|]}$,
${S^x = \frac{1}{2}  [ | \frac{3}{2}\rangle  \langle{- \frac{3}{2}}| +  | {- \frac{3}{2}}\rangle\langle   \frac{3}{2}|] }$,
and ${S^y = \frac{1}{2}  [-i | \frac{3}{2}\rangle  \langle{- \frac{3}{2}}| + i | {- \frac{3}{2}}\rangle\langle   \frac{3}{2}|] }$,
where $S^y$ transforms as a magnetic octupole and 
the remaining components transform as a magnetic dipole. 
The generic spin model for the DO doublets
on the pyrochlore lattice is given as~\cite{PhysRevLett.112.167203,PhysRevB.95.041106}, 
\begin{eqnarray}
H &=& \sum_{\langle ij \rangle} 
J_y S_i^y S_j^y + J_x S_i^x S_j^x + J_z S_i^z S_j^z +J_{xz} (S_i^x S_j^z + S_i^z S_j^x)
\nonumber \\
&& - \sum_i {\boldsymbol h}\cdot   (S^z_i \, \hat{z}_i), 
\label{eq1}
\end{eqnarray}
where only the nearest-neighbor coupling is considered and a Zeeman coupling
is introduced. Here due to the microscopic reason, only the $S^z$ component
couples linearly with the external magnetic field, and $\hat{z}_i$ is the local $z$ direction (see Table.~\ref{tab1}).
The octupolar U(1) spin liquid is realized in the regime with the dominant and 
antiferromagnetic $J_y$ coupling, and is a distinct symmetry enriched U(1) 
spin liquid in 3D~\cite{PhysRevLett.112.167203,PhysRevB.95.041106}. As a 3D U(1) spin liquid, 
the octupolar U(1) spin liquid 
shares its universal properties such as the emergent quantum electrodynamics 
for a 3D U(1) spin liquid with the elementary excitations like the gapless photon, 
the ``magnetic monopole'' and the spinon. The way 
how the emergent electric and magnetic fields in the octupolar U(1) spin liquid
are related to the physical spin operators, however, is fundamentally different 
from the conventional dipolar U(1) spin liquid. In the octupolar U(1) spin liquid,
the emergent electric field is related to the octupole moment. 
More substantially, because the $S^z$ operator flips the $S^y$ component 
and creates the spinon-antispinon pair in the octupolar U(1) spin liquid, 
the selective Zeeman coupling in Eq.~\eqref{eq1} implies that, 
the magnetic measurement such as the inelastic neutron scattering 
measurement would only probe the spinon matter~\cite{PhysRevB.95.041106}. 

 

There were several attempts~\cite{PhysRevLett.129.097202,PhysRevB.106.064427,PhysRevResearch.2.023253,PhysRevB.102.104408,Bhardwaj_2022} including our earlier ones with other authors~\cite{PhysRevB.95.041106,PhysRevResearch.2.013334,Gao_2019,Gao_2022} 
that tried to establish 
the connection between the possible spin liquid state in the Ce-pyrochlores
and the spin liquid ground states for the DO doublets. 
The representative 
proposals were based on the spectroscopic ones or the thermodynamic fitting,
and are closely related to the existing experiments. 
The octupolar U(1) spin liquid seems promising for these Ce-pyrochlores. 
In this work, we attempt to make a clear distinction between the thermodynamic 
and the spectroscopic measurements for the octupolar U(1) spin liquid with 
the DO doublets on the pyrochlore magnets. We were partly
inspired by the recent results of the thermodynamics on the Ce-pyrochlore in Ref.~\onlinecite{Bhardwaj_2022}. 
Due to the chosen question, this work here will begin with the physical explanation 
and experimental clarification, and then be supplemented with more concrete calculation.

Historically, the distinction in the thermodynamic 
and the spectroscopic measurements was crucial 
in the seminal example of fractional quantum Hall effects and represents 
an important outcome of the quantum number fractionalization 
and the deconfinement~\cite{Xiao:803748}. For the ${\nu=1/3}$ Laughlin state,  
the specific heat has an activated behavior at low temperatures, 
and the activation gap in the thermodynamics is essentially   
the gap of the Laughlin quasiparticle, $\Delta_l$. This is because 
the low-temperature specific heat probes the gas of the deconfined 
Laughlin quasiparticles. In contrast, the electron spectral function also give a gap,
and this electron spectral gap $\Delta_e$ is simply three times of the gap of the 
Laughlin quasiparticle with ${\Delta_e =3 \Delta_l}$. This remarkable result occurs 
because one single electron excites three Laughlin quasiparticles 
 in the $\nu=1/3$ fractional quantum Hall liquid. 
Thus, an analogous behavior for the octupolar U(1) spin liquid
would be quite useful to identify the deconfined and fractionalized quasiparticles. 
This seems to be difficult because there are three different quasiparticles 
in the octupolar U(1) spin liquid and all of them may contribute to the measurements
whereas there only exists one type of Laughlin quasiparticle in the ${\nu=1/3}$ Laughlin state.
Remarkably, as we will elaborate in this work, the unique properties 
of the octupolar U(1) spin liquid and the DO doublet may potentially 
make this goal feasible experimentally. In addition, we discuss the 
response of the system in the magnetic fields in both weak field regime and 
the strong field regime. 

The remaining parts of the paper are organized as follows. In Sec.~\ref{sec2}, 
we explain the thermodynamic properties, especially the spinon gap in the 
local magnetic susceptibility.
In Sec.~\ref{sec3}, we turn to the spectroscopic measurements and emphasize
the behavior of two-spinon gaps in the magnetic fields along different directions. 
In Sec.~\ref{sec4}, we explain the magnon gaps and the magnon presence for the strong field limit,
and especially the emergent properties for the field along the 110 direction. 
In Sec.~\ref{sec5}, we discuss various experimental aspects and the application
to other systems. 


\section{Thermodynamics}
\label{sec2}

We start with the simplest thermodynamic property, i.e. the specific heat. 
The specific heat probes all the excitations that include the gapless gauge photon, 
the gapped ``magnetic monopoles'', and the gapped spinons. 
While the gapless gauge photon gives a $T^3$ specific heat 
with an anomalously large coefficient~\cite{PhysRevLett.108.037202}, 
this behavior appears at very low temperatures and is very hard to be observed 
experimentally. The gapped ``magnetic monopole'' has a similar energy scale 
as the gauge photon and contributes to the specific heat in an activated fashion. 
The gapped spinon appears at a higher energy, and also contributes 
to the specific heat in an activated fashion. 
Due to the deconfined nature of the ``magnetic monopole''
and the spinon, the thermal activation of them at very low temperatures 
creates a dilute gas of both.  
The spinon experiences the ``magnetic monopole'' as the flux, 
and there exist statistical interactions between them. 
Moreover, the spinons (``magnetic monopoles'') interact with themselves
with the Coulomb interaction due to the emergent electric (magnetic) charge 
that are carried by them. Because the thermal activation leads to a soup
of three different excitations, it might be difficult to separately identify
each contribution. Nevertheless, it is still important to emphasize that, 
in the dilute gas limit where the quasiparticle interactions can be neglected, 
the specific heat then probes the properties of individual particles. 
Thus, the activated contribution from the spinon gas simply reveals 
the gap and the density of states of the individual spinon. 
This is also true for the ``magnetic monopoles''. 
If one is able to extract the activation gaps from the specific heat, 
the gaps would be associated with a single spinon or ``magnetic monopole''. 
The above discussion leads to the following qualitative expression for 
the low-temperature specific heat 
\begin{equation}
C_v (T) \sim f_p T^3  + f_s e^{-\Delta_s/T} +f_m e^{-\Delta_m/T},
\label{eq2}
\end{equation}
where $\Delta_s$ and $\Delta_m$ are the gap of single spinon and single 
``magnetic monopole'', respectively. 
Here $f_p, f_s$ and $f_m$ are the prefactors that depend on 
the properties like the density of states for these quasiparticle excitations.

\begin{table}[b]
\begin{tabular}{lcccc}
\hline\hline
$\mu$ & 0 & 1 & 2 & 3
\vspace{1mm}
 \\
\hline
$\hat{z}_{\mu}$ & $\frac{1}{\sqrt{3}} [111]$ & $\frac{1}{\sqrt{3}} [1\bar{1}\bar{1}]$ & 
$\frac{1}{\sqrt{3}} [\bar{1}1\bar{1}]$  & $\frac{1}{\sqrt{3}} [ \bar{1}\bar{1} 1] $
\vspace{2mm}
\\
$\hat{e}_{\mu}$ &  $\frac{1}{4} [111]$ & $\frac{1}{4} [1\bar{1}\bar{1}]$ & 
$\frac{1}{4} [\bar{1}1\bar{1}]$  & $\frac{1}{4} [ \bar{1}\bar{1} 1] $
\vspace{2mm}\\
\hline\hline
\end{tabular}
\caption{The local $\hat{z}$ directions of the four sublattices,
and the four neighboring vectors that connect the 
tetrahedral centers. }
\label{tab1}
\end{table}

 We now turn our attention to the other common thermodynamic quantity, i.e.
 the magnetic susceptibility. From Eq.~\eqref{eq1}, only the $S^z$ 
 component is linearly coupled to the magnetic field. 
 Since $S^z$ creates the spinon pair, then the low-temperature magnetic susceptibility
 is probing the magnetic properties of the spinon gas in the dilute spinon regime. 
Thus, the thermal activation of spinons would show up in the magnetic susceptibility. 
On the other hand, for the spin-orbit-coupled systems where the continuous spin rotational
symmetry is absent, the total magnetization is not a good quantum number to characterize 
the many-body quantum states~\cite{PhysRevB.78.094403}. 
Thus, the magnetic susceptibility in the zero temperature
limit is nonzero. To summarize, one expects that the low-temperature 
magnetic susceptibility behaves like 
\begin{equation}
\chi \sim \chi_0 + g_s e^{-\Delta_s/T} ,
\label{eq3}
\end{equation}
where $\chi_0$ is the constant zero-temperature magnetic susceptibility, 
and $g_s$ is a prefactor that depends on the properties like the spinon density of states. 
Experimentally, the above behavior is more suitable to be measured from the Knight
shift in the NMR or $\mu$SR measurement. 
Since the NMR or $\mu$SR measurements are often performed at small 
and finite magnetic fields, the relation in Eq.~\eqref{eq3} should be modified with the field. 
As long as the system remains in the spin liquid state, the gap $\Delta_s$ is the spinon
gap except that this gap would be modified by the field, and $\chi_0$ is replaced by the one in the field. 
The behavior of the spinon gap in the field will be explained in the later sections~\cite{PhysRevB.95.041106}. 
The generic expression of Eq.~\eqref{eq3} should still hold at finite and small fields.

 \section{Spectroscopic measurements}
 \label{sec3}
 
How about the spectroscopic measurements? The spectroscopic measurement
such as the inelastic neutron scattering is a data-rich measurement, and provides 
a lot of information about the excitation spectra including the dispersion, the density
of states and the spectral intensity. Again, due to the selective coupling 
with the $S^z$ components, only the $S^z$-$S^z$ dynamic spin structure factor
is measured in the spectroscopic measurements~\cite{PhysRevB.95.041106,PhysRevResearch.2.013334}. 
Therefore, the inelastic neutron scattering and the $1/T_1$ spin lattice relaxation time
measurements detect the two-spinon continuum. Thus,
the spectral gap in the spectroscopic measurements directly reveals
the gap of the two-spinon continuum. This spectral gap, 
$\Delta_{c}$, is twice of the spinon gap $\Delta_s$ in Eq.~\eqref{eq2} 
and Eq.~\eqref{eq3} with
\begin{eqnarray}
\Delta_{c} = 2 \Delta_s. 
\end{eqnarray}
This factor of 2 is an important consequence of fractionalization.

After the above qualitative physical explanation and reasoning, 
we turn to the more concrete calculation. In fact, the two-spinon 
continuum of the octupolar U(1) spin liquid
with $0$-flux or $\pi$-flux was actually explored in our early work, 
but the questions that were addressed were about the spectral periodicity 
associated with the symmetry enrichment and the translation symmetry 
fractionalization and the field-modulated spinon dispersion and Anderson-Higgs' 
 transition~\cite{PhysRevB.95.041106,PhysRevResearch.2.013334}. 
 So far, the existing experiments have not explored the continuous excitation from the 
 perspective of the spectral periodicity. This was partly due to the experimental resolution.

 The single spinon gap and excitation in the thermodynamic quantities 
 were not paid much attention.
Neither was the clear physical distinction between the thermodynamics and 
the spectroscopy. It is a bit illuminating for us to compare with the fractional 
quantum Hall context and to clarify the relation between the single spinon gap from 
 the thermodynamics and the spectral gap of the two-spinon continuum needs to be clarified.

\begin{figure}[t]
\centering
\includegraphics[width=8cm]{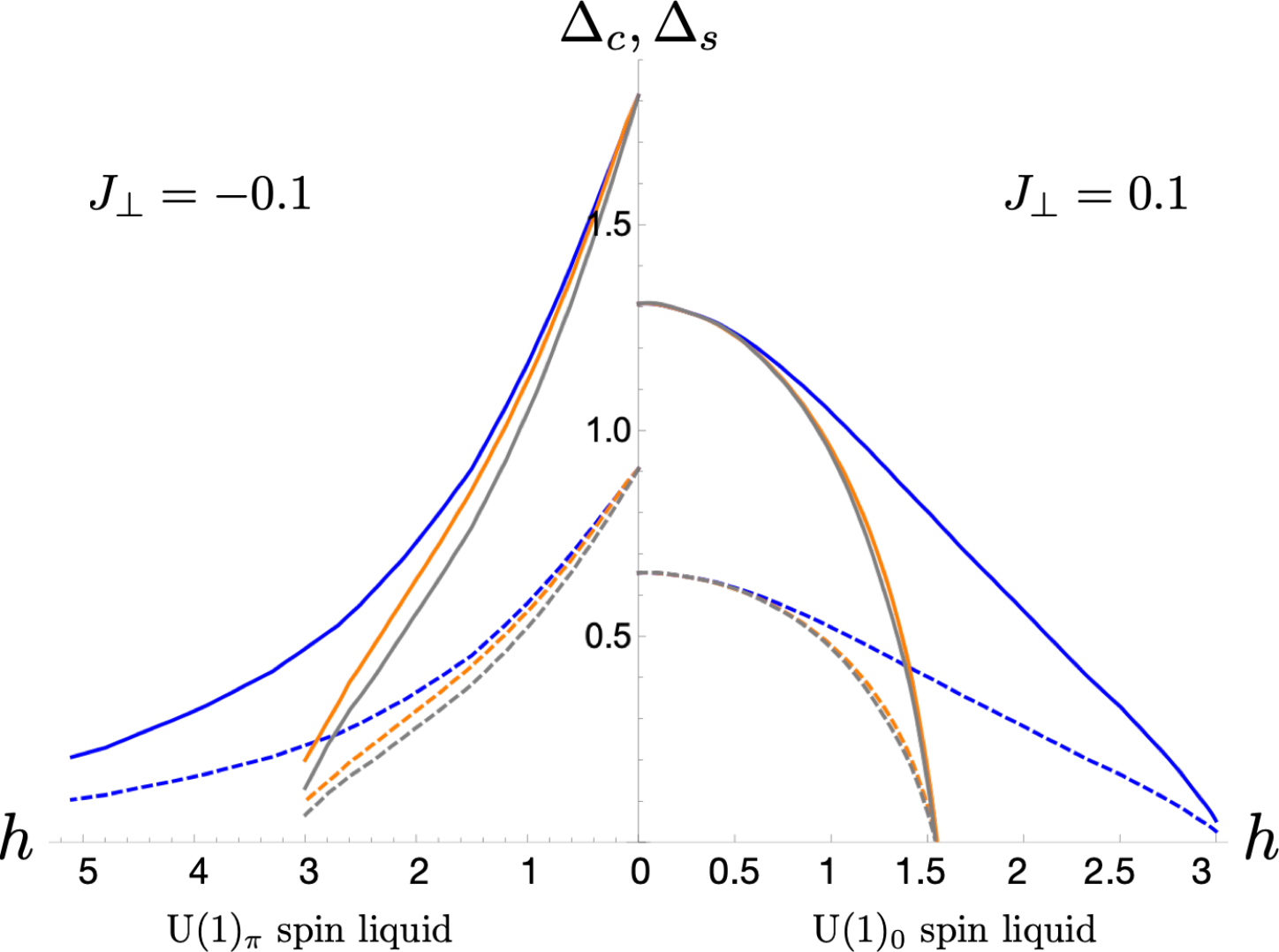}
\caption{ 
The evolution of the spinon gaps with the magnetic fields
of different directions for both 
U(1)$_0$ and U(1)$_{\pi}$ spin liquids. 
The left (right) panel is for the U(1)$_0$ (U(1)$_\pi$) spin liquid with ${J_{\perp}=-0.1 }$
(${J_{\perp}=0.1}$). 
The solid (dashed) curve is the gap $\Delta_c$ ($\Delta_s$) of the two-spinon continuum (single spinon). 
The gray, orange and blue curves correspond to the magnetic field along 001, 111, and 110 directions, respectively. 
We have set ${J_y =1}$. 
}
\label{fig1}
\end{figure}
 
We begin with the octupolar U(1) spin liquid with the $0$-flux for the spinon matter
and refer it as octupolar U(1)$_0$ spin liquid. 
This state is realized in the regime where there exists a dominant
and antiferromagnetic $J_y$ coupling and the other exchanges are 
unfrustrated. For the sake of the completeness, we here formulate 
the spinon dynamics from the microscopic spin model in this regime~\cite{PhysRevB.95.041106,PhysRevResearch.2.013334}. 
This formulation can be found in previous works~\cite{PhysRevLett.108.037202}, 
but addresses new questions in this work. 
For the purpose here, we consider a reduced model from Eq.~\eqref{eq1} in this regime,
\begin{eqnarray}
H_{\text{OSL}} = \sum_{\langle ij \rangle} J_y S^y_i S^y_j - J_{\perp} ( S^+_i S^-_j + h.c.) 
- \sum_i {\boldsymbol h}\cdot   (S^z_i \, \hat{z}_i),
\end{eqnarray}
where ${S^{\pm}_i \equiv S^z_i \pm i S^x_i}$. 
This octupolar U(1)$_0$ spin liquid occurs for a predominant $J_y$ coupling
and unfrustrated transverse exchanges.  To reveal the 
spinon-gauge coupling, we implement the well-known parton-gauge construction with
\begin{eqnarray}
S^y_i= s^y_{\boldsymbol{r} \boldsymbol{r}'}, 
\quad 
S^+_i = \Phi^{\dagger}_{\boldsymbol r} 
\Phi^{}_{\boldsymbol r'} s^+_{  {\boldsymbol r}  {\boldsymbol r} '}, 
\end{eqnarray}
 where $\Phi_{\boldsymbol r}^\dagger$ ($\Phi_{\boldsymbol r}^{}$) 
creates (annihilates) a spinon at the 
tetrahedral center ${\boldsymbol r}$. These tetrahedral centers actually form a diamond lattice
and are labeled by ${\boldsymbol r}, {\boldsymbol r}'$ to distinguish from the pyrochlore sites
$i,j$. 
We have
\begin{eqnarray}
H_{\text{OSL}} &  =& \sum_{\boldsymbol r}  \frac{J_y}{2} Q^2_{\boldsymbol r} 
 - \sum_{ \langle {\boldsymbol r} {\boldsymbol r}' \rangle  } 
 \sum_{\langle  {\boldsymbol r}' {\boldsymbol r}''  \rangle}  J_{\perp} 
 \Phi^{\dagger}_{\boldsymbol r}   \Phi^{}_{\boldsymbol r'} 
 s^+_{{\boldsymbol r} {\boldsymbol r} ''} s^-_{{\boldsymbol r}'' {\boldsymbol r}'    }
 \nonumber \\
 && \quad\quad \quad\quad  -  \sum_{\langle {\boldsymbol r}{\boldsymbol r}' \rangle} \frac{1}{2} 
 ({\boldsymbol h} \cdot {\hat{z}_i})
 (\Phi^\dagger_{\boldsymbol r} \Phi^{}_{\boldsymbol r'} s^+_{ {\boldsymbol r}{\boldsymbol r'} }  + h.c.).
 \label{eqQSL}
\end{eqnarray}
 Here the operators $s^y, s^{\pm}$ refer to the emergent U(1) gauge fields 
 such that ${s^y_{{\boldsymbol r} {\boldsymbol r}'} \simeq E_{{\boldsymbol r} {\boldsymbol r}'}}$
 and
 $ {s^+_{{\boldsymbol r} {\boldsymbol r}'} \simeq 1/2 e^{i A_{{\boldsymbol r} {\boldsymbol r}'}} }$, 
 and $Q_{\boldsymbol r} = \eta_{\boldsymbol r} \sum_{\mu} S^y_{{\boldsymbol r} {\boldsymbol r} + \eta_{\boldsymbol r} 
 {\boldsymbol e}_{\mu} }$ is imposed to enforce the physical Hilbert space, where 
 ${\eta_{\boldsymbol r} = +1 (-1)}$ for the I (II) sublattice and ${\boldsymbol e}_{\mu}$'s 
 are the first neighboring vectors on the diamond lattice. Moreover, $Q_{\boldsymbol r}$ 
 measures the electric charge density at ${\boldsymbol r}$ and thus satisfies 
 ${[\Phi_{\boldsymbol r}, Q_{\boldsymbol r'}] = \Phi_{\boldsymbol r} \delta_{\boldsymbol{r} {\boldsymbol r}'} }$,
 $[\Phi_{\boldsymbol r}^\dagger, Q_{\boldsymbol r'}] = - \Phi^\dagger_{\boldsymbol r} \delta_{\boldsymbol{r} {\boldsymbol r}'} $.

\begin{figure}[t]
\centering
\includegraphics[width=8cm]{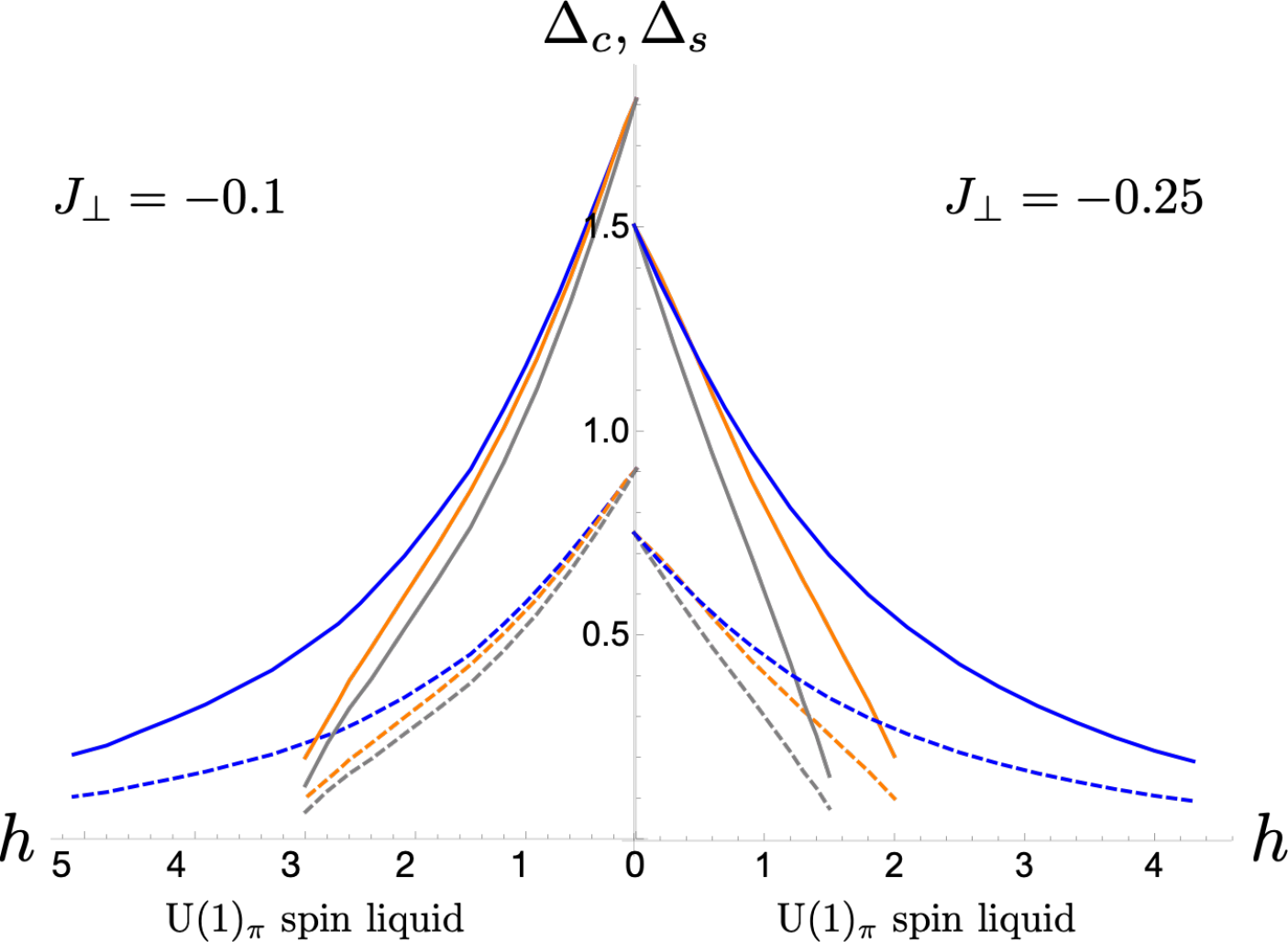}
\caption{ 
The evolution of the spinon gaps with the magnetic fields
of different directions for  
U(1)$_{\pi}$ spin liquid with different transverse couplings.
The left (right) panel is for the U(1)$_{\pi}$ spin liquid with $J_{\perp}=-0.1 $
($J_{\perp}=-0.25$). 
The solid (dashed) curve is the gap $\Delta_c$ ($\Delta_s$) of the two-spinon continuum (single spinon). 
The gray, orange and blue curves correspond to the magnetic field along 001, 111, and 110 directions, respectively. 
We have set $J_y =1$. 
}
\label{fig2}
\end{figure}
 
For the octupolar U(1)$_0$ spin liquid, one sets ${A_{{\boldsymbol r} {\boldsymbol r}'} = 0}$.
The Hamiltonian in Eq.~\eqref{eqQSL} is solved with the rotor approximation ${\Phi_{\boldsymbol r} = e^{- i \phi_{\boldsymbol r} }}$
where ${ |\Phi_{\boldsymbol r}| =1}$ and $[\phi_{\boldsymbol r}, Q_{\boldsymbol r'}] = i \delta_{ {\boldsymbol r}{\boldsymbol r}'}$. 
In the actual calculation, a Lagrangian multiplier is introduced to impose the unimodular constraint for $\Phi_{\boldsymbol r}$. 
  The spinon dispersions are obtained as $\epsilon_{\mu} ({\boldsymbol k})$ where $\mu$ is the band index. 
  For the octupolar U(1)$_0$ spin liquid, the translation symmetry is not fractionalized in a nontrivial way, and thus 
  the spinon band number is essentially the sublattice number of the diamond lattice. Because  
  the external magnetic field induces the inter-sublattice spinon tunneling in Eq.~\eqref{eqQSL}, 
  the spinon bands are not degenerate. 
  For the octupolar U(1) spin liquid when ${J_{\perp}<0}$ is 
  frustrated, a background $\pi$ flux is experienced for the spinons. 
  The U(1) spin liquid in this regime is referred as the octupolar U(1)$_{\pi}$ spin liquid. 
  One fixes the gauge to take care of the background 
  $\pi$ flux for the spinons, and the translation symmetry is fractionalized in a nontrivial way for the spinons. 
  The spinons now have four bands. Throughout the calculation, we use the $\pi$ flux 
  for the spinons. It is likely that, around the points where the spinon gap closes,
  the $\pi$ flux state may give away to other competing states.

\begin{figure}[t]
\centering
\includegraphics[width=6cm]{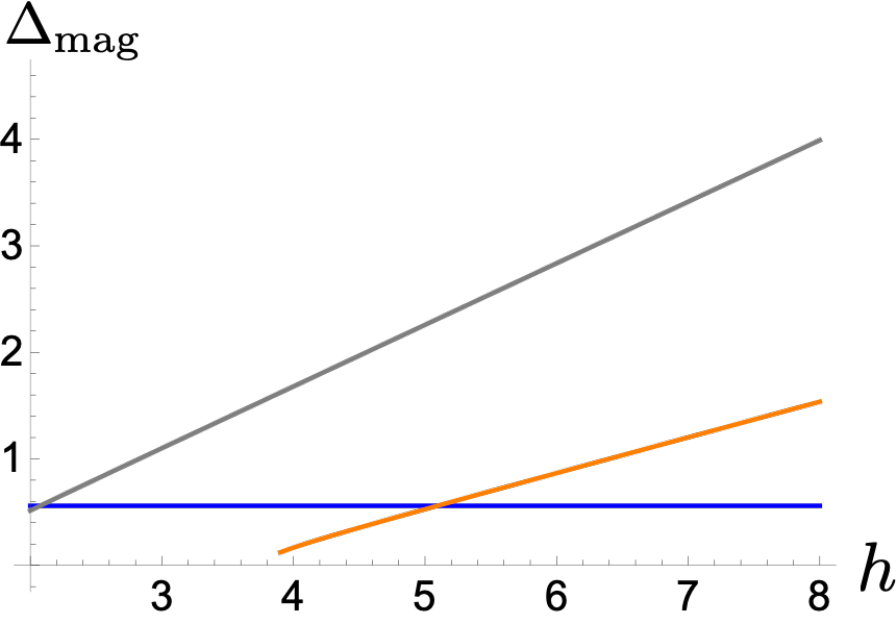}
\caption{The evolution of the magnon gap with the magnetic field
in the strong field limit for the field polarized states. The gray, 
orange and blue curves correspond to the field along 001, 111, and 110 directions,
 respectively.  
The constant behavior of the magnon gap with the 110 field
 is explained in the main text. 
 In the plot, we choose ${J_y=1, J_x=0.75, J_z=0.25}$ 
 that is compatible with the results in Ref.~\onlinecite{Bhardwaj_2022}. 
}
\label{fig3}
\end{figure}

 We label the spinon dispersions as $\epsilon_{\mu} ({\boldsymbol k})$ where 
 ${\mu =1,2}$ [$1,2,3,4$] for the octupolar U(1)$_{0}$ [U(1)$_{\pi}$] spin liquid. 
 The single spinon gap $\Delta_s$ and the gap $\Delta_c$ for the two-spinon continuum
 are given as 
\begin{eqnarray}
\Delta_s &=& \text{Min}[\epsilon_{\mu} ({\boldsymbol k})] , \label{deltas} \\
\Delta_c &=& \text{Min}[\epsilon_{\mu} ({\boldsymbol k}_1 ) + \epsilon_{\nu}( {\boldsymbol k}_2)  ] \label{delatc} .
\end{eqnarray}
The above expressions work for both octupolar U(1)$_{0}$ and U(1)$_{\pi}$ spin liquids. 
For the octupolar U(1)$_{\pi}$ spin liquid, an offset momentum from the gauge fixing
can be added to Eq.~\eqref{delatc}. Nevertheless, because the spinon energy is invariant 
under the translation by this momentum, so Eq.~\eqref{delatc} is sufficient. 

We perform an involved calculation of the spinon gaps with the self-consistent 
gauge mean-field theory for a wide range of magnetic fields and exchange couplings,
and obtain the dependence of the spinon gaps on these parameters. 
In Fig.~\ref{fig1}, we plot the comparison of the spinon gaps 
for the octupolar U(1)$_{0}$ and U(1)$_{\pi}$ spin liquids (with
the same $|J_{\perp}|$) in the magnetic fields along different directions.  
Several key features can be obtained from the plots in Fig.~\ref{fig1}. 
Firstly, the gap of the spinon continuum is twice of the single spinon gap.
Secondly, the spin liquid state is more robust against the fields for the U(1)$_{\pi}$ spin liquid. 
This is clearly expected as the $\pi$ flux is due to the more frustrated transverse exchange. 
Thirdly, the spin liquid states are more stable when the field is along 110 direction. This is because 
the field only couples two sublattices of the pyrochlore lattice and thus is less effective compared to
the fields along other directions. 
Fourthly, the spinon gap monotonically {\sl decreases} as the field increases
for all directions. Although the field does not suppress the spin liquid ground state immediately, 
the influence of the spinon dispersion
and the spinon gap is quite remarkable and is a visible effect experimentally. 
The monotonically decreasing spinon gap contrasts strongly with the 
magnon gap behavior for the field polarized states that we will explain later. 

In Fig.~\ref{fig2}, we further compare the spinon gaps 
for the octupolar U(1)$_{\pi}$ spin liquid with different $J_{\perp}$'s. 
According to Ref.~\onlinecite{Bhardwaj_2022}, ${J_{\perp} =-0.25J_y}$ is expected 
to be more relevant for Ce$_2$Zr$_2$O$_7$. There are not much
qualitatively differences for different $J_{\perp}$'s except that 
the gap is a bit smaller for large $|J_{\perp}|$.

\section{Magnon gaps in the strong field regime}
\label{sec4}

In the strong field regime, the spin liquid state will be destroyed, and the system 
becomes a field-induced polarized state. 
For the field polarized states in the strong field limit, the states for the 001 and 111 fields 
are quite clear. The 001 field induces a ${{\bf q}=0}$ two-in two-out dipolar spin ice state
in the $S^z$ components. The 111 field induces a ${{\bf q}=0}$ one-in three-out spin state 
in the $S^z$ components. These two states have been partially analyzed 
in Ref.~\onlinecite{PhysRevResearch.2.013334}. 
Here we compute the magnon spectra and find that the magnon gaps 
for these two states {\sl increase} monotonically with the field strength (see Fig.~\ref{fig3}). 
Due to the polarization effect on the $S^z$ component, the spectral weight 
of the magnons for these polarized states in the neutron scattering experiments 
should be strongly suppressed. The field-polarized state along the 110 field, 
however, is quite different from the above two field directions.

\subsection{Hidden octupolar order in the 110 field} 

It turns out that, the field-polarized state for the 110 field is quite special in the octupolar 
spin ice regime and has not been studied previously. 
It is shown below that, 
the remaining unpolarized sublattices experience a $\mathbb{Z}_2$ symmetry breaking
and induce an internal hidden order in the octupole moment at ${T=0}$. This spontaneous 
hidden octupolar order turns out to be a {\sl unique} property for the octupolar spin ice. 
Although the $\mathbb{Z}_2$ symmetry breaking on such chains along the unpolarized sublattices 
are independent, by assuming a ${{\bf q}=0}$ state, we find that, 
the magnon gap remains constant (see Fig.~\ref{fig3})
and is mainly controlled by the octupolar coupling $J_y$.

Here we discuss the field polarized states in the strong field limit. 
For the magnetic fields along 001 and 111 directions, since all
the sublattices are coupled to the field with the Zeeman coupling,
all the sublattices will be polarized by the field along their 
corresponding $+\hat{z}$ or $-\hat{z}$ directions. 
In particular, the 001 field induces 
a uniform two-in two-out spin state in the $S^z$ components
and the 111 field induces a three-in one-out spin state in the $S^z$ components.
The magnon gaps of these two polarized states grow as the field strength
is increased (see Fig.~\ref{fig3}). 
For the 110 field, however, only two sublattices are coupled 
to the field, and the other two sublattices do not directly experience the field. 

\begin{figure}[t]
\centering
\includegraphics[width=6cm]{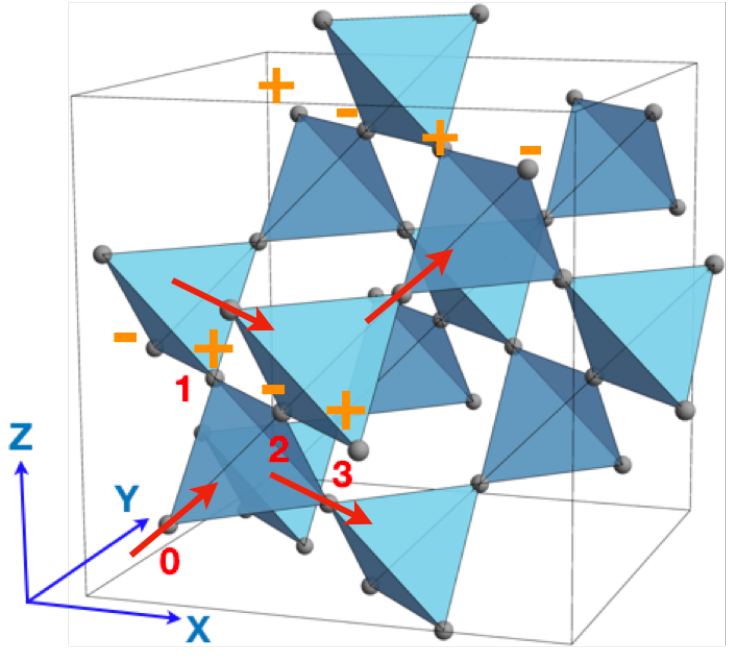}
\caption{ 
The field induced hidden ordered 
state for the 110 field. In the plot, the red arrows indicate the $S^z$ configuration
while the ``$+$'' and ``$-$'' refer to the $S^y$ configuration due to the additional symmetry 
breaking on the unpolarized sublattices.  }
\label{fig4}
\end{figure}

In Table.~\ref{tab1}, we have listed the local $\hat{z}$ directions for each sublattice. 
Under the 110 field, the 0th sublattice is polarized in the $+\hat{z}$ direction,
and the 3rd sublattice is polarized in the $-\hat{z}$ direction. 
This is depicted as the red arrows in Fig.~\ref{fig4}. 
The 1st and 2nd sublattices are decoupled from the magnetic field. 
Moreover, the internal exchange field on the 1st and 2nd sublattices from the polarized 
$\langle S^z \rangle $ of the 0th and 3rd sublattices is canceled, thus the 1st and 2nd sublattices 
are effectively decoupled from the 0th and 3rd sublattices at the classical or mean-field sense. 
The magnetic state of the 1st and 2nd sublattices will then be determined by the 
residual interaction. The 1st and 2nd sublattices in fact form multiple chains along 
$1\bar{1}0$ directions. 
Since we are considering the system in the octupolar spin ice regime with a predominant 
$J_y$ coupling, the interactions on these chains are dominated by the Ising interaction
on the octupole $S^y$ components, and these chains are effectively decoupled from each other 
in the classical sense.  
Even with the field, the Hamiltonian of the system is still invariant under the $\mathbb{Z}_2$ transformation,
\begin{eqnarray}
\mathbb{Z}_2: \quad\quad S^y \rightarrow -S^y. 
\end{eqnarray}
The predominant antiferromagnetic intra-chain $J_y$ coupling breaks this $\mathbb{Z}_2$
symmetry by creating an antiferromagnetic octupolar $\langle S^y \rangle$ order along the chains (see Fig.~\ref{fig4}).
Remarkably, the exchange field from the $\langle S^y \rangle$ configuration on the 1st and 2nd sublattices
is canceled on the 0th and 3rd sublattices. Thus, the antiferromagnetic octupolar order has no direct impact 
on the 0th and 3rd sublattices at the mean-field sense. 
As a result, the inter-chain coupling via the intermediate 1st and 2nd sublattices is absent at the mean-field level. 
The antiferromagnetic octupolar order on each chain is then independent and has a 2-fold degeneracy 
on each chain.

To summarize, the antiferromagnetic octupolar $\langle S^y \rangle$ order on the $1\bar{1}0$ chains, that is 
actually a hidden order, arises from the predominant $J_y$ coupling in the octupolar spin ice regime
after polarizing the other two sublattices in $S^z$ antiferromagnetically.  
Thus, it is one {\sl unique} property of the octupolar spin ice.

\subsection{The magnon excitation for the 110 field}

What about the magnon excitations in the field-induced states for the 110 field? 
Apparently, there are two types of magnon-like excitations. 
One is from the flipping of $\langle S^z \rangle$ on the 0th and 3rd sublattices. 
The energy gap of these magnons is governed by the magnetic field and would grow with the magnetic field. 
The other type of magnons is from the flipping of $\langle S^y \rangle$ on the 
1st and 2nd sublattices. 
The energy gap of these magnons is governed by the $J^y$ coupling and 
seems insensitive to the magnetic field. 
To demonstrate this, we perform an explicit spin-wave calculation.

We consider a ${{\bf q} =0}$ state in Fig.~\ref{fig4} such that the octupolar order on the 
effectively decoupled $1\bar{1}0$ chains are related by lattice translations. 
The reason that we choose this simple state is for our convenience to 
do the spin-wave calculation, and the mean-field analysis 
does not have the interchain correlation of the antiferromagnetic octupolar order. 
We here adopt the thermodynamic result in Ref.~\onlinecite{Bhardwaj_2022} 
and consider the model spin Hamiltonian in Eq.~\eqref{eq1} with
${J_y=1}, {J_x=0.75}, {J_z=0.25}, {J_{xz}=0}$. In the linear spin-wave theory, we express
the spin operators via the Holstein-Primakoff transformation. 
For the 0th and 3rd sublattices, we have
\begin{eqnarray}
i \in 0 , \quad S_i^z &=&\frac{1}{2} - a^\dagger_i a^{}_i , \\
S_i^x &=& \frac{1}{2} (a^{}_i  + a^{\dagger}_i ) , \\
S_i^y &=& \frac{1}{2i} (a^{}_i  - a^{\dagger}_i) , \\
i \in 3, \quad S_i^z &=&-\frac{1}{2} + b^\dagger_i b^{}_i , \\
S_i^x &=& \frac{1}{2} (b^{}_i  + b^{\dagger}_i ) , \\
S_i^y &=& \frac{1}{2i} (b^{\dagger}_i  - b^{}_i) , 
\end{eqnarray}
and for the 1st and 2nd sublattices, we have
\begin{eqnarray}
i \in 1, \quad S_i^y &=& \frac{1}{2} - c^\dagger_i c^{}_i ,    \\
 S_i^z &=& \frac{1}{2} (c^{}_i + c^{\dagger}_i ),                  \\
 S_i^x &=& \frac{1}{2i} ( c^{}_i - c^{\dagger}_i ),                 \\
 i \in 2, \quad S_i^y &=& -\frac{1}{2} + d^\dagger_i d^{}_i ,\\ 
S_i^z &=& \frac{1}{2} (d^{}_i + d^{\dagger}_i )  , \\ 
S_i^x &=& \frac{1}{2i} ( d^{\dagger}_i - d^{}_i ) .
\end{eqnarray}
The linear spin-wave Hamiltonian is then given as 
\begin{eqnarray}
H_{\text{sw}} = \sum_{\boldsymbol k} \sum_{\mu\nu} 
\Psi_{{\boldsymbol k},\mu }^{\dagger} M_{\mu\nu} ({\boldsymbol k}) \Psi_{{\boldsymbol k},\nu }^{ }, 
\end{eqnarray}
where 
$\Psi_{{\boldsymbol k},\nu }^{ } = ( a_{\boldsymbol k}, b_{\boldsymbol k}, c_{\boldsymbol k}, d_{\boldsymbol k},
a_{-{\boldsymbol k}}^\dagger, b_{-{\boldsymbol k}}^\dagger, c^\dagger_{-{\boldsymbol k}}, d^\dagger_{-{\boldsymbol k}}  )^T$,
and $M_{\mu\nu} ({\boldsymbol k})$ is a $8\times 8$ matrix with
\begin{eqnarray}
M ({\boldsymbol k}) = 
\left[
\begin{array}{cc}
A_{\boldsymbol k} &  B_{\boldsymbol k} \\
B_{\boldsymbol k}^{\dagger} & A^{\ast}_{-{\boldsymbol k}} 
\end{array}
\right].
\end{eqnarray}
Here both $A_{\boldsymbol k} $ and $B_{\boldsymbol k} $ are 
$4\times 4$ matrices with
\begin{widetext}
\begin{eqnarray}
A_{\boldsymbol k} & = &  
\left[
\begin{array}{cccc}
\frac{1}{2} (J_z  + h) & \frac{1}{4} (J_x - J_y) \cos  ( {{\boldsymbol k} \cdot {\boldsymbol e}_{03} } )
& \frac{1}{4i}J_x \cos ( {\boldsymbol k} \cdot  {\boldsymbol e}_{01}  )
& -\frac{1}{4i}  J_x \cos ( {\boldsymbol k} \cdot {\boldsymbol e}_{02}   ) 
\vspace{2mm}
\\
 \frac{1}{4} (J_x - J_y) \cos  ( {{\boldsymbol k} \cdot {\boldsymbol e}_{03} } ) & \frac{1}{2} (J_z  + h)
 & \frac{1}{4i}J_x \cos ( {\boldsymbol k} \cdot  {\boldsymbol e}_{13}  ) 
 & -\frac{1}{4i}  J_x \cos ( {\boldsymbol k} \cdot {\boldsymbol e}_{23}   ) 
 \vspace{2mm}
 \\
- \frac{1}{4i}J_x \cos ( {\boldsymbol k} \cdot  {\boldsymbol e}_{01}  ) 
& -  \frac{1}{4i}J_x \cos ( {\boldsymbol k} \cdot  {\boldsymbol e}_{13}  ) 
& \frac{1}{2} J_y 
& \frac{1}{4} (J_z - J_x ) \cos (   {\boldsymbol k} \cdot  {\boldsymbol e}_{12}  )  
\vspace{2mm}
\\
\frac{1}{4i}  J_x \cos ( {\boldsymbol k} \cdot {\boldsymbol e}_{02}   ) 
&
\frac{1}{4i}  J_x \cos ( {\boldsymbol k} \cdot {\boldsymbol e}_{23}   ) 
& 
 \frac{1}{4} (J_z - J_x ) \cos (   {\boldsymbol k} \cdot  {\boldsymbol e}_{12}  )  
 &
 \frac{1}{2} J_y
\end{array}
\right], \\
\vspace{2mm}
B_{\boldsymbol k} & = &   
\left[
\begin{array}{cccc}
0 & \frac{1}{4} (J_x + J_y) \cos ({\boldsymbol k} \cdot {\boldsymbol e}_{03}) 
& -\frac{1}{4i} J_x \cos ({\boldsymbol k} \cdot {\boldsymbol e}_{01}) 
& \frac{1}{4i} J_x \cos ({\boldsymbol k} \cdot {\boldsymbol e}_{02}) 
\vspace{2mm} 
\\
\frac{1}{4} (J_x + J_y) \cos ({\boldsymbol k} \cdot {\boldsymbol e}_{03}) 
& 0
& - \frac{1}{4i} J_x \cos ({\boldsymbol k} \cdot {\boldsymbol e}_{13}) 
&  \frac{1}{4i} J_x \cos ({\boldsymbol k} \cdot {\boldsymbol e}_{23}) 
\vspace{2mm}
\\
-\frac{1}{4i} J_x \cos ({\boldsymbol k} \cdot {\boldsymbol e}_{01}) 
&
- \frac{1}{4i} J_x \cos ({\boldsymbol k} \cdot {\boldsymbol e}_{13}) 
& 0
& \frac{1}{4i} (J_x+J_z) \cos ({\boldsymbol k} \cdot {\boldsymbol e}_{12}) 
\vspace{2mm}
\\
\frac{1}{4i} J_x \cos ({\boldsymbol k} \cdot {\boldsymbol e}_{02}) &
 \frac{1}{4i} J_x \cos ({\boldsymbol k} \cdot {\boldsymbol e}_{23}) 
&
\frac{1}{4i} (J_x+J_z) \cos ({\boldsymbol k} \cdot {\boldsymbol e}_{12}) 
&
0
\end{array}
\right],
\end{eqnarray}
\end{widetext}
where ${{\boldsymbol e}_{\mu\nu} = \frac{1}{2} ( {\boldsymbol e}_{\mu}-{\boldsymbol e}_{\nu} ) }$. 
The spin-wave spectrum is obtained by solving $H_{\text{sw}}$ 
with the bosonic Bogoliubov transformation. 
The lowest magnon gap is found to be
\begin{eqnarray}
\Delta_{\text{mag}} = [(J_y - J_x) (J_y + J_z)]^{1/2}.
\end{eqnarray}
 In the large $J_y$ limit, ${\Delta_{\text{mag}} \simeq J_y}$ and is 
essentially governed by the intra-chain coupling
along the $1\bar{1}0$ chains.

How about the spectral intensity of the magnon excitations in the inelastic neutron scattering? 
For the polarized states with the 001 and 111 fields, 
the $S^z$ components are fully polarized in the strong field limit. 
The spin operators that create magnon excitations are mainly $S^x$ and $S^y$ components
that do not couple to the external fields at the linear level. Thus, the spectral
intensity is expected to be gradually suppressed as the field is increased. 
The field-induced octupolar ordered state for the 110 field, however, behaves quite differently. 
Along the $1\bar{1}0$ chains, the system orders antiferromagnetically in $S^y$, and 
the $S^z$ component that couples to the neutron spins works as a spin flipping 
operator for $S^y$ and creates the magnon excitation. The spectral intensity of these intra-chain
magnons should be persistent even in the strong field limit and is a signature  
of hidden order in the system~\cite{PhysRevB.94.201114,PhysRevB.98.045119,Shen_2019,PhysRevResearch.2.043013}.




\section{Discussion}
\label{sec5}

\subsection{More discussion about spinon gaps}

The multiplicity relation between the thermodynamic gap from the Knight shift and the spectral 
gap from the inelastic neutron scattering or $1/T_1$ spin lattice relaxation time measurement is a sharp consequence 
of deconfinement and fractionalization. All the results in this work are about the energy gaps and did not 
invoke any information about the crystal momentum. Although high-quality single crystal samples may be ideal,
the observation of these results actually does not require the single crystal samples. 
Thus, the Knight shifts and the spin lattice relaxation $1/T_1$ of the NMR and $\mu$SR 
measurements as well as the inelastic neutron scattering can all be 
performed with the powder or polycrystalline samples, except that the 
analysis of the NMR and $\mu$SR results require the local structural 
information. 

Since the NMR and $\mu$SR measurements are often performed with the magnetic field, 
so it is better not to apply very strong magnetic fields to destroy the spin liquid. 
From the previous experiences, the spin liquid is more stable when the field is in the 110 direction.
The proposed octupolar U(1)$_{\pi}$ spin liquid
for the Ce pyrochlores is more stable in the field compared to U(1)$_{0}$ spin liquid. These
all suggest the experimental feasibility for field measurements~\cite{smith2023quantum}. 
Moreover, the spinon gap decreases with the magnetic field while the magnon gap of the 
field-polarized state does not decrease with the field. This can be a sharp and important
result to distinguish the spinon from the magnon of the candidate states.

\subsection{Discussion of field-induced competing states}

Here we give more discussion about the field-induced states for a generic magnetic field where
every sublattice is coupled to field. 
In the octupolar U(1) spin liquid, the background gauge 
flux is mainly controlled by the transverse exchange $J_{\perp}$ 
term via 3rd order perturbation theory that generates the 
ring exchange to control the background gauge flux~\cite{PhysRevB.69.064404}.   
The Zeeman coupling with the magnetic field alone enters the ring exchange via 
6th order perturbation. Thus the spin liquid state is more stable with the field perturbation
compared to the $J_{\perp}$ exchange. Nevertheless,
it is likely that, the $J_{\perp}$ exchange and the field favor different background flux.
This flux frustration can happen in the octupolar U(1)$_{\pi}$ spin liquid with the field. 
Moreover, a generic magnetic field alone would favor a uniform spin state. 
The proximate ordered state out of the octupolar U(1)$_{\pi}$ spin liquid would 
enlarge the unit cell and break the translation symmetry. 
Thus the uniform polarized state is not connected with the octupolar U(1)$_{\pi}$ spin liquid 
with a continuous or nearly continuous transition. Either the system experiences
a first order phase transition, or the system goes through an intermediate state. 
The upper-field phase boundary of the octupolar U(1)$_{\pi}$ spin liquid in 
Fig.~\ref{fig2} and the lower-field phase boundary of the polarized states in Fig.~\ref{fig3}
are not precisely known in our work. 
We have used the gauge mean-field theory for the system inside the spin liquid state,
and have used the conventional mean-field theory with spin-wave theory for the 
field-induced states. These are rather different mean-field approaches, so one cannot
simply compare the mean-field energies from them and thus cannot
obtain a phase diagram by combining these different approaches. 

For the special field along 110 direction, without the transverse exchange and other interactions, this Zeeman coupling alone,
that only acts on two sublattices, cannot generate the quantum tunneling between
 distinct spin ice configurations of the predominant $J_y$ Ising coupling. Only with other exchange interactions, the system 
 can generate quantum tunneling between
 distinct spin ice configurations. For other high symmetry field directions, the Zeeman coupling
 alone could generate the quantum tunneling between
 distinct spin ice configurations.  In our actual calculation for the 
 spinon properties~\cite{PhysRevB.95.041106}, we simply fix the background U(1)
 gauge for the spinons according to the sign of the transverse exchange.

\subsection{Discussion about other applications}

One can extend the discussion to the pyrochlore U(1) spin liquid with non-Kramers doublets. 
This can be potentially relevant for the pyrochlore magnets with the Pr$^{3+}$ and Tb$^{3+}$ ions
that include for example Pr$_2$Zr$_2$O$_7$, Pr$_2$Hf$_2$O$_7$ and Tb$_2$Ti$_2$O$_7$~\cite{Sibille_2018,PhysRevB.94.024436,PhysRevB.94.165153,PhysRevLett.118.107206,PhysRevLett.98.157204}.
Here, the spinon does not obviously show up in the magnetic measurement. 
Instead, it is the gauge photon and ``magnetic monopole'' that contribute~\cite{PhysRevB.94.205107,PhysRevB.96.195127}.
Separating the contribution of the photon and the ``magnetic monopole'' can be an obstacle. 
On the broad context, the experimental identification of fractionalized excitations
and their nontrivial symmetry enrichment in spin liquids
is an unresolved question. On the thermodynamic side, there has been some efforts
 to detect the fractional spin quantum number of the spinon in the kagom\'{e} spin liquid
 candidate~\cite{PhysRevB.92.220102}. More generally, the distinction of 
 thermodynamics and different spectroscopy can be quite useful in fulfilling this task.


\section*{Acknowledgements}

We are indebted to Jiawei Mei for many illuminating discussion
and Bruce Gaulin's group for some early communications. 
This work is supported by the Ministry of Science and 
Technology of China with Grants No.2021YFA1400300,
 the National Science Foundation 
of China with Grant No.~92065203,
and by the Research Grants Council of Hong Kong with Collaborative Research Fund C7012-21GF.

\bibliography{refs}

\end{document}